\documentclass[12pt,a4paper]{article}%

\usepackage{amsmath,amssymb,amsfonts}
\usepackage{caption2}
\usepackage[]{hyperref}
\usepackage[pdftex]{color,graphicx}
\usepackage{multicol}

\numberwithin{equation}{section} 
\setcaptionmargin{1cm}
\newcommand{\hhref}[1]{\href{http://arxiv.org/abs/#1}{{\it arXiv:#1}}}
\newcommand{\spazio}{\vspace{0.3cm}}

\usepackage{color}
\definecolor{rosso}{cmyk}{0,1,1,0.4}
\definecolor{rossos}{cmyk}{0,1,1,0.55}
\definecolor{rossoc}{cmyk}{0,1,1,0.2}

\begin{document}

\begin{titlepage}
$\quad$
\vskip 1.0cm
\begin{center}
{\huge \bf \color{rossos} Supersymmetry phenomenology beyond\\[3mm] the MSSM after 5/fb of LHC data }  
\vskip 1.0cm {\large Paolo Lodone } \\[1cm]
{
   Institut de Th\'eorie des Ph\'enom\`enes Physiques, EPFL, Lausanne, Switzerland  
} \\[5mm]
\vskip 1.0cm
\end{center}

\begin{abstract}
We briefly review the status of motivated beyond-the-MSSM phenomenology in the light of the LHC searches to date.
In particular, we discuss the conceptual consequences of the exclusion bounds, of the hint for a Higgs boson at about 125 GeV, and of interpreting the excess of direct CP violation in the charm sector as a signal of New Physics.
We try to go into the various topics in a compact way while providing a relatively rich list of references, with particular attention to the most recent developments.
\footnote{Based on an invited \href{https://indico.cern.ch/contributionDisplay.py?sessionId=10&contribId=180&confId=153252}{talk} given at \href{http://indico.cern.ch/conferenceOtherViews.py?view=standard&confId=153252}{``XX International Workshop on Deep Inelastic Scattering and related subjects''}, 26-30 March 2012, Bonn, Germany.}
\end{abstract}

\tableofcontents

\end{titlepage}

\section{Introduction}

The theoretical difficulty in understanding the smallness of the Fermi scale in the Standard Model (SM) with respect to any New Physics at very high energies is the main motivation for expecting new phenomena to show up already in the energy range that is probed by the CERN LHC.
One of the most appealing solutions to this `Hierarchy Problem' is Supersymmetry (SUSY)\footnote{A reasonalbly fair `historical list of references' could be the following: for the initial activity (before 1980) \cite{Fayet:1976cr}; for softly broken Supersymmetry \cite{Dimopoulos:1981zb}\cite{Girardello:1981wz}; for minimal Supergravity (mSUGRA) \cite{Barbieri:1982eh}-\cite{Nilles:1983ge}; for gauge madiation \cite{Dine:1981za}-\cite{Giudice:1998bp}.}, to the point that it is sometimes called `the Standard Way beyond the SM'.

Since superpartners have not been found at colliders, Supersymmetry is apparently pushed more and more to higher energies and this in principle weakens its power in solving the Hierarchy Problem, or equivalently some amount of finetuning is apparently reintroduced in the determination of the electroweak scale.
While this is more and more unavoidably true at least for its minimal implementation, the Minimal Supersimmetric Standard Model (MSSM), it is also clear that non-minimal implementations or non-standard configurations can still have the chance to be natural, i.e. not finely-tuned.
It is then crucial, from a conceptual point of view, to keep an eye on these natural configurations until they are excluded since, if it were not for the Hierarchy Problem, SUSY could manifest itself at much higher energies\footnote{Other motivations for (some) SUSY particles to be light are discussed in Section \ref{sect:split}.}.

\begin{figure}[t]
\begin{center}
\includegraphics[width=0.5\textwidth]{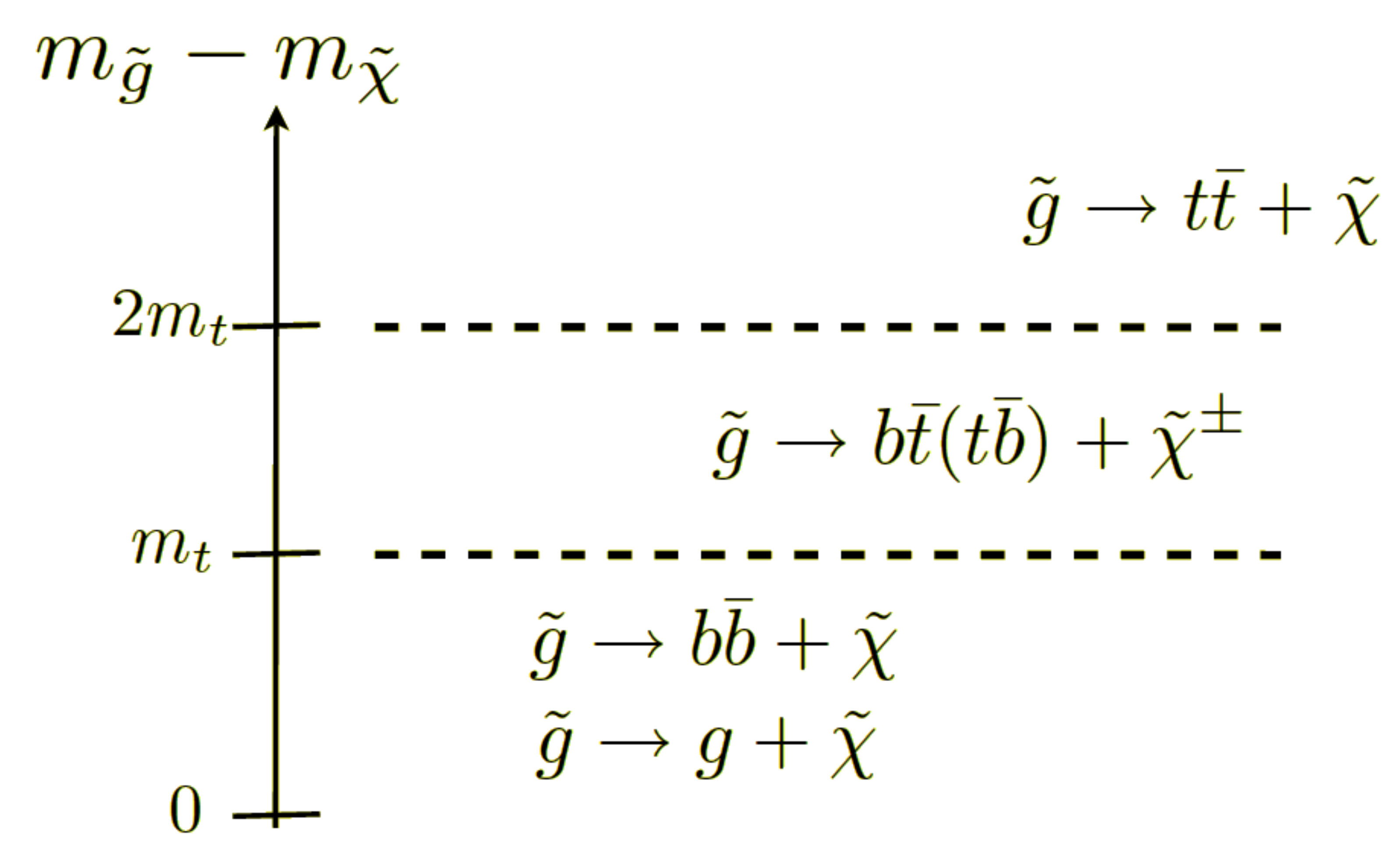} 
\caption{\small {\it A synthetic description of the relevant LHC phenomenology for Supersymmetry with a light third generation. When phase space opens up for final states involving the top quark, the channel $\tilde{g}\rightarrow b\overline{b} \chi$ is suppressed because of the smaller Yukawa coupling. See \cite{Barbieri:2009ev}\cite{talksBarbieri}. }}
\label{fig:GluinoDecays}
\end{center}
\end{figure}

The main recent experimental inputs that are relevent for our considerations are the following:
\begin{itemize}

\item A SM-like Higgs boson is now excluded from about 127 GeV up to 600 GeV \cite{Chatrchyan:2012tx}\cite{atlasHiggs}.
At the same time, there is a $2\div 3\sigma$ hint for a Higgs-like scalar close to 125 GeV.

\item Direct searches of s-particles \cite{Chatrchyan:2011zy}-\cite{atlas3genGM} have already set very strong lower bounds on their masses.
These bounds are quite model dependent, however it can be said that typically the squarks of the first two generations have to be heavier than about 1 TeV.
The case in which only the third generation is light, which is very motivated by naturalness as we shall see, is instead much less constrained, see \cite{Papucci:2011wy}\cite{Desai:2011th} for theoretical estimates.
In such a situation the gluino could still be as light as $600\div 800$ GeV, with the third generation even down to $200\div 300$ GeV, and the LHC phenomenology depends crucially on the mass difference between the gluino and the charginos/neutralinos (see Figure \ref{fig:GluinoDecays}).


\item LHCb \cite{Aaij:2011in}, and later CDF \cite{CDFcharm}, have measured with a high precision the direct CP violation in the decays of a $D$ meson into $K^+ K^-$ and $\pi^+ \pi^-$.
The world average now yields: 
\begin{equation} \label{eq:deltaAcp}
\Delta a_{CP}^{dir} = a_K^{dir}-a_{\pi}^{dir} = (- 0.67 \pm 0.16) \%
\end{equation}
which deviates by approximately 3.8$\sigma$ from the no-CP violation point, and is quite larger than the value $\approx -0.1\%$ that one would expect in the SM.

\end{itemize}

We focus on naturalness considerations in Section \ref{sect:nat}, we consider other possible issues and points of view in Section \ref{sect:other}, and then conclude in Section \ref{sect:concl}.
The main purpose of this discussion is to conceptually clarify the various possibilities in a compact way, while providing a relatively rich list of references with particular attention to the most recent studies.
We hope that this summary may be useful at least as an orientation in this subject that is now a very hot topic \cite{talksBarbieri}\cite{talkHall}\cite{talkNima}.

\section{Insisting on Naturalness} \label{sect:nat}

\subsection{The problem of finetuning}

As already said Naturalness, or the problem of finetuning\footnote{The conventional definition of finetuning is \cite{Barbieri:1987fn,Dimopoulos:1995mi}. For a modern introduction to the concept of Naturalness, see for example \cite{Luty:2005sn}; for a more extended philosophical discussion see \cite{Giudice:2008bi}; the present discussion refers to \cite{Barbieri:1996qp}\cite{Lodone:2011aa}.}, is the main theoretical motivation for physics beyond the SM at the LHC energies. It makes thus sense to discuss this point in some detail.

It is usually said that the problem stems from the `quardatic divergences' that are present in the radiative corrections to the mass of an elementary scalar, or from the fact that these corrections are quadratically sensitive to the ultraviolet cutoff of the theory.
This statement may sound strange, because it seems to depend on the way the theory is regularized: for example, where is the problem if one uses dimensional regularization instead of a sharp momentum cutoff to make the loop integrals finite?
Given the importance of the issue, let us briefly see how it can be expressed in terms of renormalized quantities: this will give us more insight in the meaning of finetuning in the supersymmetric case.
\spazio

\begin{figure}[t]
\begin{center}
\begin{tabular}{ccc}
\includegraphics[width=0.31\textwidth]{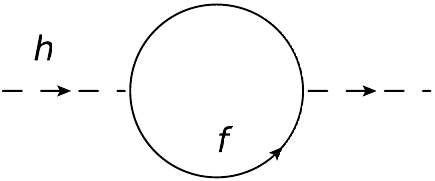} & $\,$ &
\includegraphics[width=0.58\textwidth]{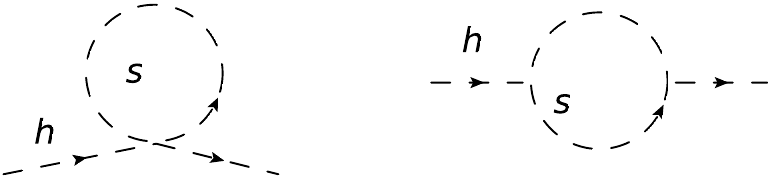}
\end{tabular}
\end{center}
\caption{\small{\it One loop corrections to the Higgs boson self energy due to fermions or scalars.}}
\label{figura:loops}
\end{figure}

Consider a heavy fermion field $f$ coupled to the Higgs boson $h$ through a Yukawa interaction:
\begin{equation} \label{eq:yukawaLagrSM}
\mathcal{L}_{Yukawa} = - y \, h \, \overline{f}_L \, f_R + h.c.
\end{equation}
and let us regularize the Higgs self energy $\Pi(p^2)$ with a cutoff $\Lambda$.
The relevant diagram is shown in Figure \ref{figura:loops} Left, and we have:
\begin{equation} \label{deltamh:fermion}
 \delta m_h^2 |_{f} = \Pi_{f}(p^2)|_{p^2=0} = \left. - \frac{y^2}{4\pi^2} \int_0^1 dx \left[ \Lambda^2 - 3 \Delta_f \log \left( \frac{\Lambda^2 + \Delta_f}{\Delta_f} \right) + ... \right]\right|_{p^2=0}
\end{equation}
where $\Delta_f=m_f^2-x(1-x)p^2$, $p$ is the momentum of the $h$ line and `$...$' are finite terms.
Let us now consider a process involving energy scales $\mu$ much higher than the value $m_h^{pole}$ of the mass of the Higgs particle when it is produced on shell (the `pole mass', that corresponds to the pole of the propagator).
If we do renormalized perturbation theory using $m_h^{pole}$, then quantum effects will give in general significant corrections to the physical observables, and it will be necessary to compute the relevant amplitudes at many orders in perturbation theory.
It is instead convenient to treat $m_h$ as a parameter of perturbation theory, defined by a renormalization prescription at the scale $\mu$, for example:
\begin{equation} \label{eq:rencond}
S^{-1}_{ren}(p^2)|_{p^2=\mu^2} = \mu^2 - m_h^2(\mu)
\end{equation}
where $S_{ren}(p^2)$ is the renormalized propagator of the scalar $h$.
As a result $m_h$ becomes an `effective mass' and starts running with the energy scale $\mu$ according to the Renormalization Group (RG) flow, like the other parameters of perturbation theory, with initial condition:
\begin{equation}
m_h(m_h^{pole}) = m_h^{pole} \, .
\end{equation}
In this way the most important quantum corrections are resummed provided that one uses the value of this `running mass' $m_h(\mu)$ to compute the mass effects in a process involving the typical momentum scale $\mu$.
In words, $m_h(\mu)$ is `the mass that minimizes the quantum corrections if we probe the theory at the energy scale $\mu$'.

Consider now a fundamental theory that describes the physics at a very high energy scale $\Lambda_{in}$:  the value of the Higgs boson mass will be derived from the fundamental parameters of the theory at this high scale. Then, from the point of view of the fundamental theory, the `initial value' of this parameter will be $m_h(\Lambda_{in})$ at the scale $\Lambda_{in}$, to be run down according to the RG flow.
Actually there is some ambiguity here, because the running mass can be defined in different ways by using renormalization conditions different from (\ref{eq:rencond}). However this does not influence the essence of the problem. In fact notice that the quadratic divergence is canceled once and for all by the counterterm (so that it does not play any physical role), and for $\mu < m_f$ the logarithm in (\ref{deltamh:fermion}) is rougly independent of $p^2$, and it is again canceled by a constant counterterm.
On the other hand for $\mu > m_f$ the correction starts being logarithmically energy dependent. This behaviour cannot be reabsorbed in the counterterms, which are polynomial functions of the momentum. The unavoidable consequence is that $m_h$ starts {\it running} according to:
\begin{equation} \label{eq:runmh:ferm}
\frac{d m_h^2 (\mu)}{d \log \mu} = - \frac{3 y^2}{4 \pi^2} m_f^2 + ... \, .
\end{equation}
where `$...$' stands for other terms that are eventually present, for example proportional to $m_h^2$ itsef.
Notice that this does not depend on having used a cutoff regularization instead of other methods, although in $m_h(\mu)$ there is some ambiguity related to the renormalization conditions that one chooses. This ambiguity however is only related to the polynomial terms in the self energy $\Pi(p^2)$ as a function of $p^2$, and it reflects the possibility of a different choice of the counterterms. The logarithmic contribution is instead completely fixed by the theory.

From the point of view of renormalized perturbative quantum field theory, the hierarchy problem stems from (\ref{eq:runmh:ferm}) and not, strictly speaking, from the quadratic divergence in (\ref{deltamh:fermion})!
Recall in fact that, for a fermion mass, the RGE is always proportional to the mass itself thanks to chiral symmetry, and thus a fermion mass `tends to remain small if it is initially small'. 
Let us think instead about what (\ref{eq:runmh:ferm}) means. Since analogous effects come from the interaction with scalar particles (Figure \ref{figura:loops} Right) or vectors, the above result shows that {\it the running mass of a scalar particle takes contributions from the mass of any particle it couples to}.
Suppose now that we look at the SM as the low energy remnant of a more complete theory, whose parameters are given at the `input scale' $\Lambda_{in}$, and we want to try to construct this theory. The question that one has to ask himself is whether to do that is easy or not.
The hierarchy problem amounts to recognize that the answer is: {\it no, it is highly nontrivial}.
In fact we have to specify the value of $m_h(\Lambda_{in})$, and then run it down to low energy in order to find the value $m_h(\Lambda_{SM})$. Let us see how {\it precise} this initial condition must be. To this end we change it by a small amount $\epsilon$, and see how the low energy theory is modified:
\begin{equation}
m_h(\Lambda_{in}) \rightarrow (1+\epsilon)m_h(\Lambda_{in}) \quad \Rightarrow \quad m_h(\Lambda_{SM}) \rightarrow (1+\Delta \, \epsilon)m_h(\Lambda_{SM}) \, .
\end{equation}
Equation (\ref{eq:runmh:ferm}) with $m_f \sim \Lambda_{in}$  is a fair way to mimic the effect of the coupling of $h$ to the high energy (or short distance) physics, and the result is:
\begin{equation} \label{eq:def:finetuning:gen}
\Delta = \frac{d \log m_h^2(\Lambda_{SM})}{d \log m_h^2(\Lambda_{in})} \sim \frac{\Lambda_{in}^2}{m_h^2(\Lambda_{SM})}
\end{equation}
which is precisely the definition of finetuning \cite{Barbieri:1987fn,Dimopoulos:1995mi}, and (\ref{eq:def:finetuning:gen}) is usually refered to as the amount of finetuning (or inverse finetuning).
The unavoidable conclusion is that, to guarantee $m_h(\Lambda_{SM}) \sim 10^2$ GeV assuming that the input scale $\Lambda_{in}$ is at least\footnote{Since we know that gravity exists and it is not included in the SM.} the Planck scale $M_{Pl} \sim 10^{19}$ GeV, the initial condition must be given at least with the precision of one part over $\Delta \sim 10^{34}$.

\spazio
As is clear from the above discussion, naturalness arguments are strongly dependent on the hypotheses that one decides to make.
For example if one assumes that the SM is the ultimate theory of nature, instead of being the low energy remnant of a more fundamental theory, then there is no problem: the Higgs boson has a mass which is exactly of the same order of the only energy scale he couples to, which is the Fermi scale $v$.
A more defendable possibility is to agree on the fact that new physics exists, and to accept the finetunig becuse of anthropic arguments\footnote{For example, the cosmological constant poses another huge unsolved naturalness problem, yet its value is close to the upper bound beyond which galaxy formation is not possible \cite{Weinberg:1987dv}. Analogously, with $m_h$ very different from the Fermi scale, we would not have atoms and again life would be impossible.}.
One can then say that, if we live in a Multiverse where different physics takes place in different Universes, our one is maybe not `natural' but it is one of the few which can allow our existence.
Another possible way to avoid the hierarchy problem is to eliminate the hierarchy, i.e. to make $M_{Pl}$ not far from the TeV scale. This is possible in the context of Large Extra Dimensions (LED) \cite{ArkaniHamed:1998rs,Antoniadis:1998ig}, in which the `volume' of the $n$ compactified extra dimensions reduces the `true' Planck scale $M_{Pl}^*$ from $M_{Pl}$ down to to:
\begin{equation}
M_{Pl} \rightarrow M_{Pl}^* = \left(\frac{M_{Pl}^{2}}{V_{(n)}}\right)^{\frac{1}{n+2}}
\end{equation}
so that it can be $M_{Pl}^* \ll M_{Pl}$ if $V_{(n)}$ is large in TeV$^{-n}$ units.

The logical choice that is behind low-energy Supersymmetry is that New Physics must respect a symmetry that protects the Higgs boson mass. Moreover the characteristic energy scale at which this symmetry (and thus this NP) manifests itself cannot be much different from the TeV scale.
To see this, let us go back to the quardatic divergences: is it wrong to talk about Naturalness in terms of them?
Suppose that the hierarchy problem is solved by means of a symmetry that protects the Higgs boson mass against large corrections. This symmetry has to be broken at low energy, and it will be restored at some higher scale $\Lambda_{NP}$. In particular at energies higher than $\Lambda_{NP}$ there will be additional particles and interactions which `symmetrize' (i.e. cancel) the contribution of SM fermions to the running of $m_h$. If we regularize the various contribuitions to the self energy with a cutoff $\Lambda$, we must have something like (\ref{deltamh:fermion}) with $m_f$ replaced by $\Lambda_{NP}$.
Then the quadratic divergence will cancel out because of the symmetry, but in general there will be finite and logarithmic terms with coefficients of the form:
\begin{equation} \label{top:cutoff:estim0}
\frac{y^2}{4 \pi^2} \Lambda_{NP}^2 \, .
\end{equation}
Since at higher energy the theory respects the symmetry, $\Lambda_{NP}$ is the largest {\it non symmetric} scale to which the Higgs boson is coupled, and thus it will replace $\Lambda_{in}$ in (\ref{eq:def:finetuning:gen}). At the end of the day we obtain, for the dominant top contribution:
\begin{equation}\label{top:cutoff:estim}
\delta m_h^2 \sim \frac{3 y_t^2}{4\pi^2} \Lambda_{NP}^2 < \Delta \times m_h^2
\quad \Rightarrow \quad
\Lambda_{NP} \lesssim \frac{m_h}{(\mbox{100 GeV})} \times \sqrt{\frac{\Delta}{10}}\times  ( 1 \mbox{ TeV})
\end{equation}
where $\Delta$ is the finetuning (or amount of cancellation) that we tolerate, as defined above.
Thus after all the usual `naive'  estimate is absolutely correct, as one would have guessed thinking in terms of effective theories.

In supersymmetry the estimates (\ref{top:cutoff:estim0}) and (\ref{top:cutoff:estim}) hold thus with:
\begin{equation} \label{eq:cutoff:estim:susy}
\Lambda_{NP} \sim \mbox{ SUSY breaking masses at the energy scale $M$.}
\end{equation}
where $M$ is the `Messenger' scale at which supersymmetry breaking is communicated to the SM.
It is the running of the relevant parameters from $M$ down to low energy that may need some amount of cancellation in order to correctly reproduce the SM.
This means that there is a {\it residual finetuning} in supersymmetric models: the point is that we now have the chance to reduce $\Delta$, as defined in (\ref{eq:def:finetuning:gen}), from $10^{34}$ down to $10^1$ [or $10^2$], which can be considered an acceptable [better than $10^{34}$ but maybe uncomfortable?] amount.

\subsection{Natural Supersymmetry}  \label{sect:natsusy}

Let us now see more precisely where is the finetuning problem in the Minimal Supersymmetric Standard Model (MSSM)\footnote{See e.g. \cite{Martin:1997ns} for an introduction and for notation and conventions.}, and how it is connected to the Higgs boson mass\footnote{For a very clear and detailed discussion of this point see \cite{Casas:2003jx}.}.
In Supersymmetry the mass of the lightest Higgs scalar is controlled by the quartic coupling of the Higgs sector. As a consequence, in the MSSM there must be a light CP-even Higgs-like scalar $h$, with the tree level relation:
\begin{equation} \label{eq:MSSM:Higgs:mass:bound}
m_h^2 \leq m_Z^2 \cos^2 (2 \beta)
\end{equation}
where $\tan \beta$ is the ratio of the vevs, $v_u / v_d$, of the two Higgs doublets $H_u$ and $H_d$.
A Higgs boson below $m_Z$ is excluded by data since a long time, so clearly extra contributions are needed.
The only possibility within the MSSM is to raise $m_h$ through radiative corrections, and the dominant contribution comes from the stop sector. Using the one loop effective potential one finds:
\begin{equation} \label{eq:mh:withradiative:stop}
m_h^2|_{1\, loop} \leq m_Z^2 \cos^2 (2\beta) + \frac{3 m_t^2}{4\pi^2 v^2} \left( \log \frac{\overline{m}_{\tilde{t}}^2}{m_t^2}   + \frac{X_t^2}{\overline{m}_{\tilde{t}}^2} \left( 1- \frac{X_t^2}{12 \overline{m}_{\tilde{t}}^2} \right)\right)
\end{equation}
where $\overline{m}_{\tilde{t}}$ it the average stop mass squared and $X_t=A_t-\mu \cot\beta$, $A_t$ being the stop A-term and $\mu$ being the superpotential mass term for the Higgs doublets. If we want to increase $m_h$ up to 125~GeV in this way, we need stop masses at least of order of some TeV.
What is the problem with that?
The problem is that, minimizing the scalar potential which leads to Electroweak Symmetry Breaking (EWSB), one finds:
\begin{eqnarray}
\frac{m_Z^2}{2} &\approx& -m_{H_u}^2 - |\mu|^2 \,  \qquad (\mbox{if }\tan\beta \gg 1). \label{eq:SMminimization:3}
\end{eqnarray}
Now, the value of $-m_{H_u}^2$ at low energy is determined by its value at the Messenger scale $M$ and by its running. For example the top system gives the one loop running:
\begin{equation}
\frac{d m^2_{H_u}}{d \log \mu} = \frac{|y_t|^2}{16\pi^2} \, \cdot \, 6  (m_{\tilde{Q}_3}^2 + m_{\tilde{u}_3}^2 + |A_t|^2) \, .
\end{equation}
Thus in general large stop masses introduce a large radiative correction on $m_{H_u}^2(100\mbox{ GeV})$ with respect to its original value $m_{H_u}^2(M) = m_{H_u}^2(100\mbox{ GeV}) - \delta m^2_{H_u}|_{rad}$.
Using the definition of finetuning (\ref{eq:def:finetuning:gen}) and fixing the amount $\Delta$ that we tolerate, we get from (\ref{eq:SMminimization:3}):
\begin{equation}
\frac{d \log m^2_Z}{d \log m^2_{H_u}(M)} \leq \Delta \quad \Rightarrow \quad \left|\delta m^2_{H_u}|_{rad}\right| \leq \Delta \cdot \frac{m_Z^2}{2}
\end{equation}
from which we have an `upper naturalness limit' on the stop masses. In the same way we obtain a `bound' on the gluino mass $m_{\tilde{g}}$, which enters in the running of all the squark masses as:
\begin{equation}
\frac{d m^2_{\tilde{u_3},\tilde{Q_3}}}{d \log \mu} = \frac{g_3^2}{16\pi^2} \, \cdot \left( -\frac{32}{3} \right) m_{\tilde{g}}^2
\end{equation}
and then into $m^2_{H_u}$ at two loops.
An upper bound on the $\mu$-term comes directly from the tree-level relation (\ref{eq:SMminimization:3}).
Putting all together one finds, neglecting $A_t$ for simplicity (see also \cite{Papucci:2011wy}):
\begin{eqnarray}
m_{\tilde{t}_{L,R}, \tilde{b}_L} & \lesssim &    \mbox{500 GeV } \, \sin\beta \, \left( \frac{3}{\log M/(\mbox{1 TeV})} \right)^{\frac{1}{2}} \left( \frac{m_h^{tree}}{\mbox{100 GeV}}\right) \left( \frac{\Delta}{10}\right)^{\frac{1}{2}}         \nonumber \\
m_{\tilde{g}} & \lesssim &    \mbox{1100 GeV } \, \sin\beta \, \left( \frac{3}{\log M/(\mbox{1 TeV})} \right)^{\frac{1}{2}} \left( \frac{m_h^{tree}}{\mbox{100 GeV}}\right) \left( \frac{\Delta}{10}\right)^{\frac{1}{2}}        \label{eq:natbounds} \\
\mu &\lesssim & \mbox{250 GeV } \left( \frac{m_h^{tree}}{\mbox{100 GeV}}\right) \left( \frac{\Delta}{10}\right)^{\frac{1}{2}}  \nonumber
\end{eqnarray}
where $m_h^{tree}$ is the lightest Higgs boson mass at tree level.
In the above discussion we kept only the bounds coming from the larger couplings, which are the top Yukawa $y_t$ and the strong gauge coupling $g_3$. The bounds on all the other superpartners are weaker, as we discuss below.

\begin{figure}[t]
\begin{center}
\includegraphics[width=1.0\textwidth]{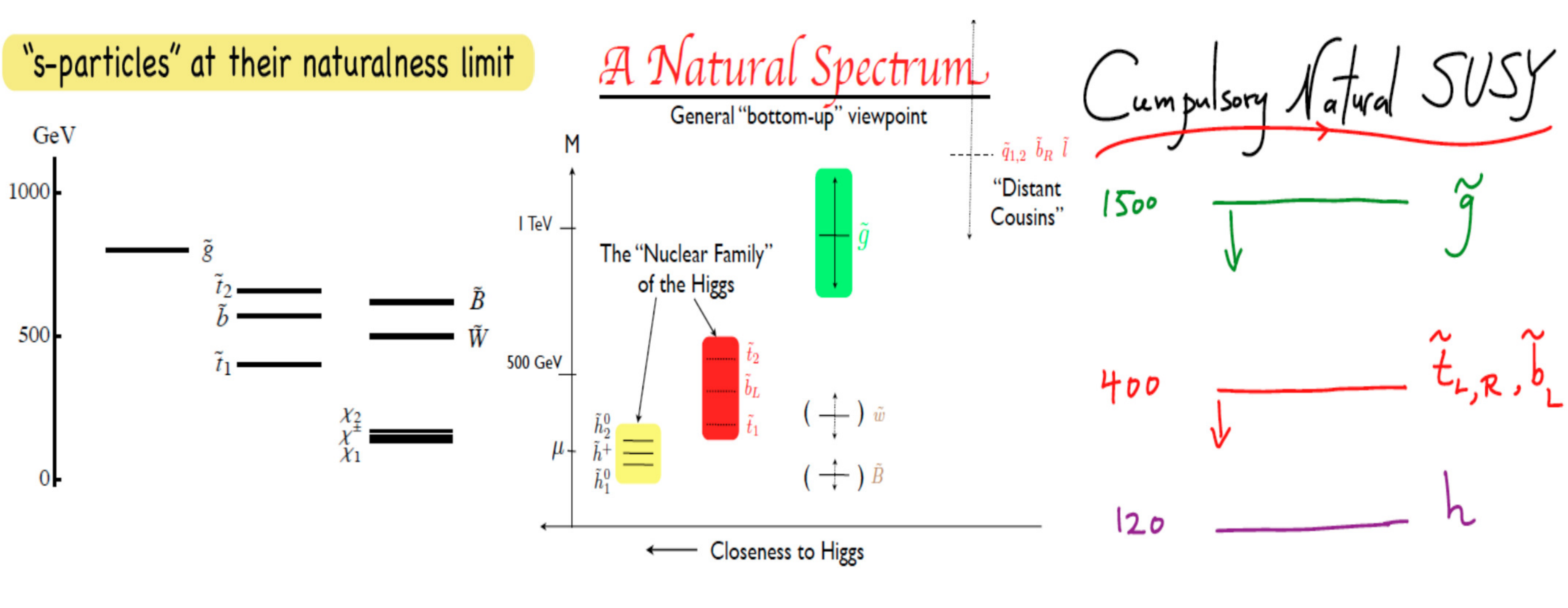} 
\caption{\small {\it Examples of a natural supersymmetrics spectrum, the other s-particles are heavier. Left: taken from \cite{Barbieri:2009ev} and \cite{talksBarbieri}. Center: taken from \cite{talkHall}. Right: taken from \cite{talkNima}.}}
\label{fig:NatSpectrum}
\end{center}
\end{figure}

This natural supersymmetric spectrum, in which the conditions (\ref{eq:natbounds}) are satisfied while the other superpartners are heavier, has been considered since a long time \cite{Dine:1990jd}-\cite{Barbieri:1997tu} as an alternative to the more conventional scenario with almost degenerate s-particles, also in connection with the `Supersymmetric Flavor Problem' and the `Supersymmetric CP problem' since heavier sfermions of the first two generations can help in satisfying the flavor and CP bounds.
This configuration, called `More Minimal Supersymmetric Standard Model', `Hierarchical Sfermions' or `Non-Standard Supersymmetric Spectrum' (or in other ways) has received considerable attention also relatively recently \cite{Barbieri:2009ev},\cite{Giudice:2008uk}-\cite{Barbieri:2011fc}, more or less in connection with Flavor and CP issues.
If one saturates (\ref{eq:natbounds}) with the choice $\Delta= 10$ and eventually assumes a relatively low scale of mediation of SUSY-braking, then one is led to consider configurations like those depicted in Figure \ref{fig:NatSpectrum}.

\spazio
What can we say about naturalness in Supersymmetry in light of the recent data?
As discussed in the Introduction, the first 5 fb$^{-1}$ of LHC data gave us at least three important messages:
\begin{enumerate}
\item The squarks of the first two generations, if there, must have masses $\gtrsim 1$ TeV.
\item The third generation is excluded only op to $\sim 300$ GeV.
\item There is a hint for the Higgs boson at about 125 GeV (with enhanced di-photon rate).
\end{enumerate}
Now, since so far the LEP bound $m_h > 114.4$ GeV was often considered one of the main problems for naturalness in Supersymmetry, one may think that the main message is the third one and that $m_h \sim 125$ GeV rules out naturalness completely.
This is absolutely not the case!
A correct statement may be that the {\it natural pure MSSM} starts being in trouble\footnote{See e.g. \cite{Hall:2011aa}-\cite{Ghilencea:2012gz} for recent discussions.}, since even with a large A-term one needs a stop mass of at least 1 TeV in equation (\ref{eq:mh:withradiative:stop}), which is too large to be natural\footnote{It is not a problem to have an enhanced di-photon rate, e.g. with a light stau \cite{Carena:2011aa}.}.

In fact it is well known that a heavy stop is not the only way to raise the lightest Higgs boson mass in Supersymmetry: there are a lot of ways to do that!
The most straightforward one is probably the Next to Minimal Supersymmetric Standard Model (NMSSM)\footnote{See \cite{Ellwanger:2009dp} for a review. See also \cite{Ross:2011xv} for a general discussion of finetuning.}, in which one adds a gauge singlet superfield $S$ with superpotential coupling to the Higgs doublets $\lambda S H_u H_d$.
As a consequence the tree-level upper bound on the mass of the lightest scalar becomes:
\begin{equation}
m_h^2 \leq m_Z^2 \cos^2{2\beta} + \lambda^2 v^2 \sin^2{2\beta}
\quad , \quad
[\mbox{NMSSM}-\lambda\mbox{SUSY}] \, .
\label{mhlsusy}
\end{equation}
While in the NMSSM one usually keeps $\lambda \lesssim 0.7$ so that the coupling $\lambda$ does not become nonperturbative below the unification scale, in $\lambda$SUSY \cite{Barbieri:2006bg} one requires perturbativity just up to 10 TeV, and $\lambda$ can be set to be equal to 2 at low energies, so that $m_h$ can be as high as 350 GeV\footnote{Notice the connection with the idea of the `fat Higgs' \cite{Harnik:2003rs}-\cite{Delgado:2005fq}.}. Notice that, although such a heavy SM-like Higgs boson is now excluded, one can still lower its mass through singlet-doublet mixing \cite{Hall:2011aa}. In this case, in (\ref{eq:natbounds}), one has to use the expression in equation (\ref{mhlsusy}) although the physical mass is smaller at tree level, because what counts there is the quartic coupling (or `the maximum $m_h$ that you can have at tree level').

Another possibility is to increase the quartic coupling by adding new gauge interactions, under which also the standard matter fields are necessarily charged. As a result one finds, at tree level:
\begin{eqnarray}
m_h^2 &\leq& (m_Z^2 +\frac{g_x^2 v^2}{2(1+\frac{M_X^2}{2 M_\phi^2})})\cos^2{2\beta} 
\quad , \quad
[U(1) \mbox{ extension}]
\label{mhU1} \\
m_h^2 &\leq&  m_Z^2 \frac{g^{\prime 2} + \eta g^2}  {g^{\prime 2} +  g^2}    \cos^2{2\beta}, 
\quad
\eta=\frac{1 + \frac{g_I^2 M_\Sigma^2}{g^2 M_X^2}}
{1+ \frac{ M_\Sigma^2}{ M_X^2}}
\quad , \quad
[SU(2)\mbox{ extension}],
\label{mhSU2}
\end{eqnarray}
where in the first case we added an extra $U(1)$ with coupling $g_x$ (see \cite{Batra:2003nj}\cite{Lodone:2010kt} for details),  $M_X$ is the mass of the new gauge boson and $M_\phi$ is the soft breaking mass of new heavy scalars. In the second case we added an extra $SU(2)$ factor (see \cite{Lodone:2010kt}-\cite{Maloney:2004rc} for details), and the standard $SU(2)_W$ gauge group is  extended to $SU(2)_I\times SU(2)_{II}$ with couplings $g_I$ and $g_{II}$, broken down to the diagonal $SU(2)$ subgroup at a higher scale so that $g = {g_I g_{II}}/{\sqrt{g_I^2 + g_{II}^2}}$; $M_\Sigma$ is the soft breaking mass of a heavy scalar in the $(2,2)$ and $M_X$ is the mass of the quasi-degenerate heavy gauge triplet vectors.

Further possibilities are adding extra vector-like matter that contributes to the Higgs boson mass through loops (see e.g. \cite{Martin:2009bg}), or introducing non-renormalizable operators \cite{Casas:2003jx},\cite{Polonsky:2000rs}-\cite{Boudjema:2012cq} that give hard SUSY-braking corrections to the Higgs quartic coupling\footnote{See \cite{Cassel:2009ps}  for a detailed study of finetuning in this case.}.

Given this plethora of possibilities for increasing the Higgs quartic beyond the minimal model, it is clear that a Higgs boson at 125 GeV is absolutely not a problem for naturalness in Supersymmetry.
The \emph{only requirement} is truly that (\ref{eq:natbounds}) are satisfied, and this is still possible with $\Delta \lesssim 10$ provided that we abandon the idea of increasing $m_h$ through the stop loops only.
Notice that this is a very nontrivial conceptual point, although it depends on the amount of finetuning that one tolerates.
In fact a Higgs boson close to 115 GeV would have been instead an indication that after all the MSSM does not need to be extended in order to make it natural. But this is not what the LHC is telling us!

\spazio
These considerations are more or less behind many of the most recent works on phenomenological Supersymmetry, e.g. \cite{Papucci:2011wy},\cite{Desai:2011th},\cite{Hall:2011aa},\cite{Kats:2011qh}-\cite{Baer1203.5539}.
For example in \cite{Hall:2011aa} the MSSM, NMSSM and $\lambda$SUSY are compared in detail in light of the Higgs boson hint at 125 GeV, showing that one needs at least $\Delta\gtrsim 100,10\div15,5$ respectively in the three cases.
Concerning the enhanced di-photon rate that is maybe needed if there is truly the Higgs boson behind the present excess at 125 GeV, it is not difficult to obtain it through mixing effects both in $\lambda$SUSY \cite{Hall:2011aa} and in the NMSSM \cite{Ellwanger:2011aa}-\cite{Ellwanger:2012ke}.

The main point, however, is that a spectrum of the type of those in Figure \ref{fig:NatSpectrum} is really the only left out possibility for Supersymmetry to be natural, given the bounds on the squarks of the first two generations\footnote{Unless the present bounds on the first two generations are escaped... see Section \ref{sect:escaping}.}. A more `conventional' degenerate spectrum would in fact imply stops above the TeV, which {\it per se} means $\Delta\gtrsim 100$ if Higgs quartic is not much increased.
For this reason, from being called `More Minimal Supersymmetric Standard Model' or `Hierarchical Sfermions' or `Split Families' or `Non-Standard Supersymmetric Spectrum', this configuration has gained the right to be called {\bf `Natural Supersymmetry'}\footnote{For explicit examples of this recent change of name, see e.g. \cite{Papucci:2011wy}\cite{talkNima}\cite{Hall:2011aa}\cite{Allanach:2012vj}\cite{Baer1203.5539}%
.}.

\begin{figure}[t]
\begin{center}
\begin{tabular}{cc}
\includegraphics[width=0.32\textwidth]{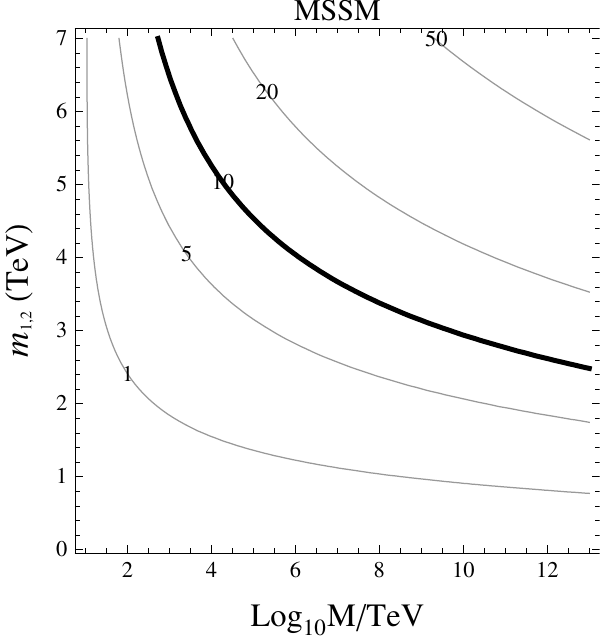} &
\includegraphics[width=0.32\textwidth]{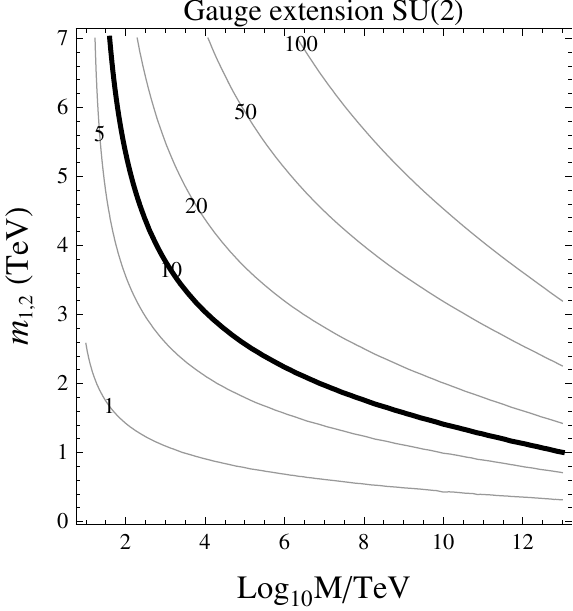} \\
\includegraphics[width=0.32\textwidth]{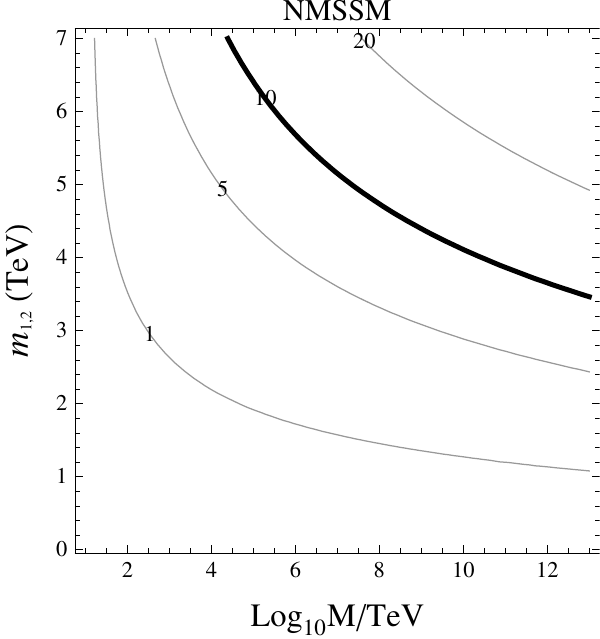} &
\includegraphics[width=0.32\textwidth]{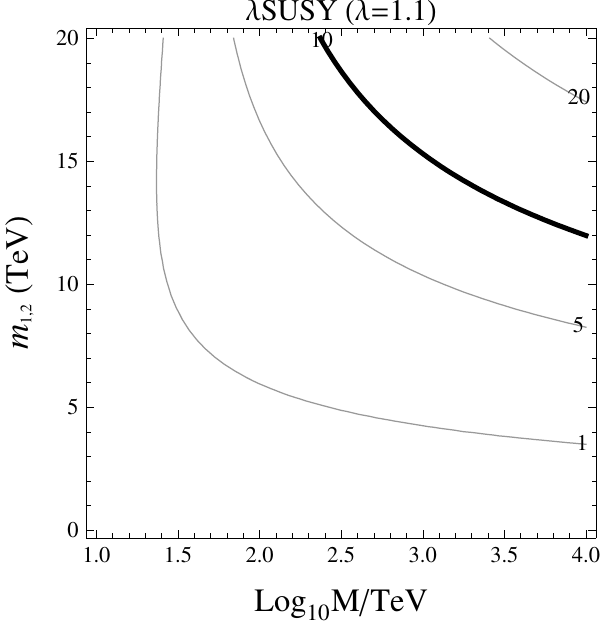} 
\end{tabular}
\end{center}
\caption{{\small \it Naturalness upper bound on the mass of the $1^{st}$ and $2^{nd}$ generation, for different values of $\Delta$, as a function of the SUSY-breaking mediation scale $M$. For $\lambda$SUSY with $\lambda v\sim200$ GeV, perturbativity is up to about $10^4$ TeV. The Higgs mass is taken to be 125 GeV (see text) and the thick line stands for $\Delta=10$.
We assume  degenerate masses at $M$ so that (\ref{eq:fayetiliopoulos}) is canceled.
In the $U(1)$ case, the bound is stronger than for $SU(2)$.}}
\label{naturalness12gen}
\end{figure}

\spazio
Finally, what about the first two generations of sfermions: how heavy can they naturally be?
To see this, recall that their soft masses enter in the RGE of the lightest Higgs boson mass (through contributions to $m^2_{H_u}$) at one loop only proportionally to the `Fayet-Iliopouolos' term:
\begin{equation} \label{eq:fayetiliopoulos}
\mbox{Tr}(Y\tilde{m}^2) = \mbox{Tr}(\tilde{m}^2_Q + \tilde{m}^2_D -2 \tilde{m}^2_U -\tilde{m}^2_L +\tilde{m}^2_E) \, .
\end{equation}
To be conservative, let us assume that this term vanishes at the scale $M$ where the renormalization group flow starts, which can happen if there is a vertical degeneracy\footnote{For simplicity we assume total degeneracy; in principle it is enough to have it within $SU(5)$ multiplets.}.
The effect arises then at the two-loop level, and in the MSSM with large $\tan\beta$ it reads:
\begin{equation}
\frac{d m_h^2}{d \log{\mu}} = \frac{48}{(16\pi^2)^2}(g^4 + \frac{5}{9} (g^{\prime})^ 4) m_{1,2}^2
\quad , \quad
[\mbox{MSSM}],
\end{equation}
where $m_{1,2}^2$ is a common mass for the first two generations.
The same expression is true also for the case of the NMSSM and $\lambda$SUSY, taking into account that for low $\tan\beta$ also the radiative corrections to $m^2_{H_d}$ come into play.
In the case of gauge extensions, on the contrary, there are additional terms due to the fact that the standard matter fields are charged under the new gauge groups. For large $\tan\beta$, so that $m_h$ is maximized given the value of the new gauge couplings, one finds:
\begin{eqnarray}
\frac{d m_h^2}{d \log{\mu}} &=& \frac{48}{(16\pi^2)^2}(g^4 + \frac{5}{9} (g^{\prime})^ 4 + \frac{7}{6}g^4_x) m_{1,2}^2
\quad , \quad
[U(1)\mbox{ extension}],
\label{U1run} \\
\frac{d m_h^2}{d \log{\mu}} &=& \frac{48}{(16\pi^2)^2}(g_I^4 + \frac{5}{9} (g^{\prime})^ 4) m_{1,2}^2
\quad , \quad
[SU(2)\mbox{ extension}],
\label{SU2run}
\end{eqnarray}

The corresponding naturalness bounds are shown in Figure \ref{naturalness12gen}.
The main features of these plots are easily understood \cite{Barbieri:2010pd}.
The known result for the MSSM is that with $\Delta\sim 10$ one needs $m_{1,2}\lesssim2\div 3$ TeV if $M$ is as high as the unification scale \cite{Dimopoulos:1995mi}, while it can be $m_{1,2}^2\sim 5\div 7$ TeV if $M$ is not beyond $10^2\div 10^3$ TeV.
The main effect is a rescaling factor (125 GeV)$/ m_Z$ in the NMSSM where there is no large mixing (neglecting the radiative corrections to $m_h$), and $\lambda v / m_Z$ in the case of $\lambda$SUSY since what counts is the quartic coupling of the Higgs sector no matter if there are mixings that reduce the actual lightest Higgs boson mass.
On the contrary in the case of gauge extensions, even if $m_h$ is increased at tree level, the new gauge interactions introduce additional quantum corrections that make the bound stronger.

\begin{table}[t]
\begin{center}
\begin{tabular}{r|c|c|c|c|c|}
  Model & $m_h=125$ GeV & $\tilde{t}$ & $R(h\rightarrow \gamma\gamma)$& tuning \\ \hline \hline
MSSM & difficult & $\stackrel{\mbox{above 1 TeV}}{\mbox{to increase } m_h}$ &  light $\tilde{\tau}$ & $\Delta \gtrsim 100$\\ \hline 
NMSSM & easy & $\stackrel{\mbox{can be light}}{\Rightarrow \mbox{ natural}}$ &   $\stackrel{\mbox{singl-doubl}}{\mbox{mixing}}$ & $\Delta \gtrsim 10\div 15$\\ \hline 
\bf{$\lambda$SUSY} & $\stackrel{\mbox{\bf{sing-doub mix}}}{\mbox{\bf{to lower it}}}$ & $\stackrel{\mbox{\bf{can be light, but also}}}{\mbox{\bf{naturally heavier!}}}$ &  $\stackrel{\mbox{\bf{singl-doubl}}}{\mbox{\bf{mixing}}}$  & $\mathbf{\Delta \gtrsim 5}$ \\ \hline 
Gauge ext. & easy & $\stackrel{\mbox{can be light}}{\Rightarrow \mbox{ natural}}$    & $\stackrel{\mbox{presumably}}{\mbox{as in MSSM}}$  & $\stackrel{\mbox{more costr. on}}{1^{st}-2^{nd} \mbox{ gen}}$\\ \hline
$\stackrel{\mbox{Nonrenorm.}}{\mbox{operators}}$ & easy & $\stackrel{\mbox{can be light}}{\Rightarrow \mbox{ natural}}$  & $\stackrel{\mbox{presumably}}{\mbox{doable}}$ & $\stackrel{\mbox{at least}}{\Delta \gtrsim 10}$ \\ \hline
$\stackrel{\mbox{Vector-like}}{\mbox{matter}}$ & $\stackrel{\mbox{borderline}}{\mbox{(depends)}}$ & $\stackrel{\mbox{can be light}}{\Rightarrow \mbox{ natural}}$  & $\stackrel{\mbox{presumably}}{\mbox{as in MSSM}}$ & $\stackrel{\mbox{at least}}{\Delta \gtrsim 10}$\\ \hline
\multicolumn{5}{c}{ } \\
  \multicolumn{3}{c}{$\qquad$ Old} & \multicolumn{2}{c}{New} \\ \hline
$\stackrel{\mbox{Name of the}}{\mbox{scenario}}$ & \multicolumn{2}{c|}{$\stackrel{\mbox{`More Minimal SSM', `Hierarchical}}{\mbox{sfermions', `Non-Standard SUSY Spectrum',...}}$}  & \multicolumn{2}{c|}{\bf{`Natural Supersymmetry'}} \\
\hline
\end{tabular}
\end{center}
\caption{{\it A summary of what discussed in Section \ref{sect:natsusy}.}}
\label{table:naturalsusy}
\end{table}

\spazio
To summarize, Natural Supersymmetry with stops and left-handed sbottom well below the TeV and heavier first two generations stands now as the only `conventional' left out possibility for Supersymmetry to solve the Hierarchy Problem with $\Delta \lesssim 10$, or equivalently 10\% finetuning (other `less conventional' possibilities are discussed in Section \ref{sect:escaping}).

Our considerations are briefly summarized in Table \ref{table:naturalsusy}.
The main point, as already said, is that the spectrum shown in Figure \ref{fig:NatSpectrum} has now the right to be called `Natural Supersymmetry'.
Moreover, if the hint for a Higgs boson at 125 GeV is confirmed, $\lambda$SUSY with singlet-doublet mixing stands as the most natural possibility. Notice also that, due to the large increase in the Higgs quartic, in $\lambda$SUSY all the squarks are naturally allowed to be heavier by a factor of 2 or 3 (see also Figure \ref{naturalness12gen}).
Thus, for example, if in the future no stops are found below 1 TeV and/or the squarks of the first two generations are excluded up to about 5 TeV, then only $\lambda$SUSY-like configurations will have the right to be called `Natural Supersymmetry'...
In this last case, the easiest way to exclude Natural Supersymmetry might be to exclude the extended Higgs sector\footnote{See e.g. \cite{Cavicchia:2007dp}.}. We will come back on these issues in the Conclusions.

As a final remark we quote that, maybe also motivated by similar considerations, there has been recent interesting model building activity focussed on SUSY breaking with split families, or Natural Supersymmetry as we now say: see e.g. the recent idea of `Flavor Mediation' \cite{Craig:2012yd} \cite{Craig:2012di} or other possibilities \cite{Craig:2011yk}\footnote{See also \cite{Barbieri:2011ci}. }   \cite{Larsen:2012rq}-\cite{Csaki:2012fh}\footnote{See \cite{Gherghetta:2003wm}-\cite{Sundrum:2009gv} for previous related work.}, including even Natural Supersymmetry from string theory \cite{Krippendorf:2012ir}.
Moreover, since a low $M$ can help in minimizing the amount of finetuning in (\ref{eq:natbounds}) given the sparticle masses, we can also say that low-scale SUSY breaking is somehow favoured by data(+naturalness); see e.g. \cite{Draper:2011aa}\cite{Petersson:2012dp} for very recent studies.

\section{Other possibilities/issues} \label{sect:other}

\subsection{Escaping the bounds}\label{sect:escaping}

Besides taking heavier superpartners, another way to keep Supersymmetry alive is escaping the experimental bounds through less-standard or peculiar configurations.

\spazio

A first well-known possibility is R-Parity violation\footnote{See \cite{Hall:1983id}-\cite{Barbieri:1985ty} for historical references, and \cite{Barbier:2004ez} for a review.}.
Recall that, from the superfield point of view, the down-type Higgs doublet $H_d$ has exactly the same quantum numbers of the left-handed lepton doublet.
It is thus clear that the Baryon Number and the Lepton Number are not accidental symmetries of the theory at the renormalizable level, as instead is the case for the SM.
Introducing all the couplings allowed by the gauge symmetry without any suppression mechanism, one ends up with large contributions to excluded processes like proton decay.
The standard way to avoid these terms is to impose by hand an additional discrete symmetry, namely the R-Parity, whose quantum number is $R_{\phi}=(-1)^{3(B-L)+2S}$ where $B$ is the baryon number, $L$ is the lepton number, and $S$ is the spin of the field $\phi$\footnote{This is better than imposing separate Baryon and Lepton number conservation for at least two reasons: neutrino oscillations, and the fact that $B$ and $L$ are separately violated by nonperturbative effects at high energies, while $B-L$ is conserved.}.
Notice that all the SM particles have R-Parity $+1$, while all the superpartners have R-Parity $-1$.
The most striking consequence of this fact is that the Lightest Supersymmetric Particle (LSP) must be stable, and can thus be a very good Dark Matter candidate.
From the point of view of collider signatures, R-Parity leads to the crucial prediction of abundance of events with missing energy, since the LSP necessarily escapes the detector.
Most of the experimental SUSY-search analyses published so far assume in fact that the production of any supersymmetric particle is accompained by the missing energy carried away by the LSP at the end, eventually, of a decay chain. 
For the case of R-Parity violating (RPV) supersymmetry, on the contrary, very few experimental analyses have been published so far, see e.g. \cite{Jackson:2011zi} \cite{ATLAS-RParity35}, and it is clear that the present stringent bounds on the `conventional' R-Parity conserving case can be escaped, at least in part.
From the conceptual point of view, for RPV to be `natural' one needs a rationale to make the effect small enough in order not to be in trouble with proton decay and other very constrained processes.
For example a mechanism that one could use, and that is well known to be very efficient in suppressing the flavor violation that additional flavor structures can introduce in beyond-the-SM contexts, is Minimal Flavor Violation (MFV)\footnote{\cite{Chivukula:1987py,Hall:1990ac,D'Ambrosio:2002ex}; see also \cite{Isidori:2012ts} for a recent review.}.
The possibility of RPV with MFV has been studied in \cite{Nikolidakis:2007fc}\cite{Smith:2008ju} and also recently in \cite{Csaki:2011ge}\cite{Arcadi:2011ug}. See also \cite{Brust:2011tb}\cite{Allanach:2012vj} for recent discussions.

Even with conserved R-Parity, the signal of Jets plus missing energy can be significantly reduced if the mass splittings between some superpartners are small. For example, a small gluino-LSP splitting reduces the phase space for the gluino dacay: as a consequence the typical missing transverse energy and momentum coming from processes involving gluinos are reduced, leading to a reduced acceptance for given selection cuts.
For recent studies of this configuration, called `Compressed Spectrum', see e.g. \cite{LeCompte1005.4304}\cite{LeCompte:2011fh}.

A similar way to reduce the missing energy, that in some cases is even more efficient, is `Stealth SUSY' \cite{Fan:2011yu}\cite{Fan:2012jf} (see also \cite{Csaki:2012fh}).
In this case, R-Parity is conserved but the LSP of the ordinary sector decays into particles belonging to a `Stealth sector' in which the relative mass splittings within a supermultiplet are small. Moreover, R-Parity-even stealth particles can decay back into SM particles; the phase space is thus again reduced for the decay chain that terminates with the true `stealth LSP'.

Another example of possible non-standard signatures is given by Dirac gauginos, which may distort the phenomenology in a significant way \cite{Hall:1990hq}\cite{Randall:1992cq}\cite{Fox:2002bu}, see also \cite{Brust:2011tb}\cite{Heikinheimo:2011fk}\cite{Kribs:2012gx} for recent studies.

\spazio

In summary, given the negative results of the SUSY searches so far, trying to see whether the detection may have been escaped for a reason that is not simply large mass is probably being to become a very hot topic in the near future, as an alternative way in which a natural Supersymmetry could be realized.
Known possibilities are suppressing the missing energy (RPV), suppressing the `visible energy' that goes into jets (Compressed Spectrum), suppressing both (Stealth SUSY), or suppressing the production cross section by giving the gauginos a large Dirac mass (Dirac gauginos).
New phenomenological studies and ideas about the above issues would probably be very timely.

\subsection{Insisting only on DM and unification and/or strings} \label{sect:split}

Recall that the main motivations for Supersymmetry are, broadly speaking: (i) It can solve the Hierarchy problem, in a perturbative way; (ii) It gives a good Dark Matter (DM) candidate; (iii) Gauge couplings unify at high energy; (iv) It is very elegant and a necessary ingredient of string theory.
Moreover all this is done in a way that is almost automatically compatible with the EWPT.

\spazio
Retaining all the features (i)-(iv) provides a very attractive picture.
However, that Nature must be without finetuning is not a necessity (we will come back on this in the Conclusions), and it is meaningful to study what happens retaining (ii)-(iv) while abandoning the requirement (i).
It is clear that for DM one needs at least some neutralino at low energy, although its mass may be relatively heavier than the usual SUSY DM. It can be seen that the only choice that is radiatively stable is to keep all the gauginos and higgsinos around the TeV scale, while the squarks and one Higgs scalar doublet can be at a very high mass scale $\tilde{m}$. Remarkably, it turns out that this is also what is required for obtaining a precise gauge coupling unification.
This scanario is called `Split Supersymmetry' \cite{ArkaniHamed:2004fb}-\cite{Bernal:2007uv}, and the only finetuning that is needed is the (large) one that makes one of the two scalar doublets much lighter than its natural mass scale, which would be $\tilde{m}$.
Split Supersymmetry has many interesting peculiar features, whose discussion go beyond our present scopes; we refer for example to \cite{ArkaniHamed:2004yi} for a review.
To cite one of them, the gluino can be very non-standard and its dominant dacay width can be into a gluon and a gravitino, giving a single-jet signal with a distinctive energy distribution.
Alternatively, although it is not the LSP, the gluino can be very long lived if its dominant decay mode is mediated by the heavy squarks, with a variety of possible unusual signatures not only at the LHC \cite{Kilian:2004uj}-\cite{Gambino:2005eh}.
The experimental discovery of a slowly decaying gluino would be a strong indication for Split SUSY, and would also provide a way to measure the high scale $\tilde{m}$.

A more radical point of view is to retain only the motivation (iv) and to accept again a large finetuning in the value of the Fermi scale. Actually, having accepted the finetuning, string theory is a very plausible framework since it provides a huge number of vacua (the `Landscape'), and one can speculate about environmental/anthropic selection among them \cite{Agrawal:1997gf}. In any case, what one ends up with is `High-Scale Supersymmetry' \cite{Hall:2009nd}, in which all the superpartners are relegated at a very high energy and DM is plausibly made of axions.

\spazio
What can LHC data to-date say about these scenarios? An interesting feature of both models is that, assuming that there is no additional field content besides the one of the MSSM, the value of the Higgs mass is very precisely determined in terms of the parameters of the theory, in particular the scale $\tilde{m}$, with some variability depending on the boundary condition at the scale $\tilde{m}$.
By carefully performing the running and the various matchings \cite{Alves:2011ug}-\cite{arXiv:1108.6077}\cite{Arbey:2011ab}, it can be seen that a Higgs mass of 125 GeV implies $\tilde{m}\lesssim 10^5$ TeV for the case of Split SUSY, while no bound can be derived on $\tilde{m}$ in the case of High-Scale SUSY.

Finally, these models are practically designed so that the direct bounds on sparticle masses do not touch them, and in particular the second case would not be touched even by eventual strong exclusion bounds on higgsinos and gauginos.
If, on the contrary, particles resembling charginos and neutralinos are found, then it will be crucial to measure their couplings with high precision and see whether they are compatible with supersymmetry or not\footnote{For these effects see \cite{KEK-PREPRINT-93-146}-\cite{hep-ph/9803210},\cite{Giudice:2004tc},\cite{Bernal:2007uv},\cite{Giardino:2011aa}.}. To do so, a Linear Collider would be probably necessary.

\subsection{Higgs boson not at 125 GeV}

So far we have assumed that the hint for a Higgs boson at 125 GeV stands indeed for a signal, and we discussed some of the consequences for supersymmetric models.
Given the importance of the issue, however, it is clear that we cannot content ourselves with a $2\div 3\sigma$ evidence, and more data is definitely needed (as the experimental collaborations also say).
Moreover the fact that the channel in which most of the excess is seen, namely the diphoton one, prefers a rate that is \emph{larger} than in the SM is precisely what one would expect in case of a background fluctuation.

Let us than say something about the possibility (that may be disproved quite soon) that the present excess goes away and a SM-like Higgs boson is excluded up to a mass of 500-600 GeV.
The SM alone would then be excluded by the combination of Higgs searches and Electroweak Precision Tests.
In the context of Supersymmetry, the most plausible explanation would be that the CP-even scalars of the Higgs sector have a reduced production cross section times branching ratio with respect to the SM Higgs boson. Namely, what counts is the quantity:
\begin{equation}
\xi_X = \frac{\sigma(pp\rightarrow s) BR(s\rightarrow X)}{\sigma(pp\rightarrow h)_{SM} BR(h\rightarrow X)_{SM}}
\end{equation}
where $X$ is a given final state.
In particular, above $2m_Z$ the Higgs boson decay into vector bosons is the dominant channel in the SM. Looking at the exclusion plots we see that a heavy `not-so-much-SM-like Higgs boson' is still allowed by data  provided that $\xi_{VV}$ ($V=W,Z$) is smaller than $0.3\div 0.5$. This suppression is not difficult to achieve, generally speaking, in supersymmetric models \cite{Spira:1995rr}\cite{Djouadi:2005gj}, thanks to mixing effecs (see e.g. \cite{Carena:2011fc}), or top-loop effects that modify the gluon-fusion efficiency \cite{Low:2009nj}, or a depletion of $BR(s\rightarrow VV)$ due to a significant decay width into two neutralino LSP.
In the case of $\lambda$SUSY, in which such non-SM-like Higgs particles above $2m_Z$ can be achieved very easily, it can be seen that it is not implausible to have $\xi_{VV} \sim 0.1\div 0.2$ for both the CP-even scalars, and it can also happen that the next to lightest state is more easily detectable than the lightest one \cite{Lodone:2011ax}-\cite{Bertuzzo:2011aa}.

\spazio

In conclusion, strictly speaking at the moment of writing it is not excluded that the hint for a Higgs boson at 125 GeV is just a fluctuation of the background. In this case there may be one or more not-so-much-SM-like Higgs bosons with mass between 200 and 300 GeV and $\xi_{VV} \lesssim 0.3 \div 0.5$.
Waiting for more data to disprove these possibilities, we can say that if no SM-like Higgs boson is found at 125 GeV, then in reasonable models we typically expect at least one almost-SM-like such scalar below 500 GeV with\footnote{We come back in the Conclusions to the possibility that noting is seen also down to $\xi\lesssim 0.1$.} $\xi \gtrsim 0.1$. If the lightest scalar, with reduced couplings, is found to be above 200 GeV, then in the context of Supersymmetry this would be a strong indication for models with largely increased quartic coupling, like $\lambda$SUSY.
On the contrary, if a SM-like Higgs boson is indeed found at 125 GeV, then eventual additional scalars can have very low $\xi$-values.


\subsection{About direct CPV in $D$ decays}

As a last issue, let us give attention to the possibility that the recently measured direct CP-Violation in the $D$-meson decays, reported in equation (\ref{eq:deltaAcp}), is a signal of physics beyond the SM. 
Whether the experimental value can be explained within the SM or not  is still under discussion. What we can say is that it is \emph{possible} that the SM accounts for the observed effect \cite{Golden:1989qx}\cite{Brod:2011re}, 
but maybe not very easily or at least it is fair to say that \emph{New Physics is not implausible}
(see e.g. \cite{Grossman:2006jg}  -\cite{Giudice:2012qq}).

\spazio
In the NP interpretation, what one needs is a rationale to understand why the main effect comes out in a $\Delta F=1$ process without disturbing the very tightly-constrained $\Delta F=2$ ones.
Referring to \cite{Giudice:2012qq} for details, a way to understand it in the context of supersymmetric models may be that flavor violation in the squark sector comes mainly from Left-Right mixings (i.e. from the A-terms), while the Left-Left and Right-Right mixings are subleading.
The size of the experimental value can then be naturally explained in the configuration called `disoriented A-terms' , in which the size of the entries of the A-term matrices is the same of those of the corresponding Yukawa matrices but without respecting exact proportionality.

A concrete realization of this configuration that may actually be considered also independently of Supersymmetry can be \cite{RattazziTalk} to invoke the paradigm of Partial Compositeness\footnote{\cite{Kaplan:1991dc}-\cite{Contino:2006nn}. See \cite{Contino:2010rs} for an introduction and \cite{Redi:2011zi} for a recent study. See also \cite{Barbieri:2008zt}.}.
In fact, if the flavor structure is originated by this mechanism, than one expects an effective Lagrangian of the type:
\begin{equation} \label{eq:finalparametrization}
\mathcal{L}_{\Delta F}^{(eff)} = \sum c_{ij}^{ab} \,\, {\epsilon_i^{a} \epsilon_j^{b} g_{\rho} v}   \, \frac{1}{\Lambda^2} \,     (Q_{\Delta F=1})_{ij}^{ab} 
+  \sum c_{ijkl}^{abcd} \,\, {\epsilon_i^{a} \epsilon_j^{b}  \epsilon_k^{c} \epsilon_l^{d} g_{\rho}^2}      \, \frac{1}{\Lambda^2} \,      (Q_{\Delta F=2})_{ijkl}^{abcd} \, \,  
\end{equation}
while the Yukawa matrices are given by:
\begin{equation}  \label{eq:yukawas}
(Y_u)_{ij} \sim g_{\rho} \epsilon^u _i \epsilon^q_j
\quad , \quad
(Y_d)_{ij} \sim g_{\rho} \epsilon^d _i \epsilon^q_j \, ,
\end{equation}
where the $\epsilon^{q,u,d}_i$ are `suppression factors' generated by the strong dynamics.
It is then clear that $\Delta F=2 $ operators tend to be suppressed more than the $\Delta F=1 $ ones, and thus we are going in the right direction.
Moreover as discussed e.g. in \cite{Isidori:2011qw} and \cite{Giudice:2012qq}, the best candidates for producing a sizable effect in the charm sector while being consistent with the other flavor constraints are the $\Delta C=1$ chromomagnetic operators, such as:
\begin{equation} \label{eq:Q12qu}
Q_{12}^{qu} = \, \overline{u}_L \sigma^{\mu\nu} g_s G_{\mu\nu} c_R\, 
\end{equation}
together with the other one with opposite chiralities whose effect is typically relatively suppressed by a factor $m_u/m_c$.
In the notation (\ref{eq:finalparametrization}) we have, following \cite{Isidori:2011qw}\cite{Giudice:2012qq}:
\begin{equation}
\Delta a_{CP} \approx -(0.13\%) \mbox{Im}(\Delta R^{SM}) - \left(\frac{3{\textrm{TeV}}}{\Lambda}\right)^2 \mbox{Im}({c^{qu}_{12}}) \mbox{Im}(\Delta R^{NP})
\end{equation}
where Im$(\Delta R^{NP})\sim 0.2$ is the hadronic matrix element of $Q_{12}^{qu} $, and we used the fact that $\epsilon^q_1$ is expected to be $\approx \epsilon^q_2$ times the Cabibbo angle. Since a reasonable estimate can be $\Delta R^{SM}\sim 1$, in order to be compatible with the observed value (\ref{eq:deltaAcp}) one needs either an enhanced SM contribution $\Delta R^{SM}\sim 5$, or a NP contribution that corresponds in our case to:
\begin{equation} \label{eq:fixingthescale}
 \Lambda = 10 \mbox{ TeV } 
\quad , \quad
\mbox{Im}(c^{qu}_{12}) \sim 1 \, .
\end{equation}
In a supersymmetric context\footnote{This configuration, sometimes called `hierarchical wavefunctions', has been studied for example in \cite{Davidson:2007si}-\cite{Dudas:2010yh}, also in the context of Supersymmetry. See also \cite{Nelson:2000sn}\cite{Nelson:2001mq} for the original idea.} these flavor-violating operators are generated by the flavor-mixings in the squark mass matrices, that are of the form (looking at the order of magnitudes, no proportionality between matrices!):
\begin{eqnarray}
(m^2_{u,d})^{LL}_{ij}  , (m^2_{u,d})^{RR}_{ij} &\sim& \tilde{m}^2_i \delta_{ij} + m_0^2 \epsilon^{u,d,q}_i \epsilon^{u,d,q}_j  \nonumber \\
(m^2_{u,d})^{LR}_{ij} \sim (m^2_{u,d})^{RL}_{ji} &\sim&  g_{\rho} \epsilon^q_i \epsilon^{u,d}_j \left( v _{u,d}\,   A_0  + v_{d,u}\, \mu \delta_{ij} \right)  \, .  \label{eq:sqark:massmatrices}
\end{eqnarray}
Moreover the effects come at the loop level, so that (\ref{eq:fixingthescale}) can translate into:
\begin{equation} \label{eq:fixingthescale2}
 \tilde{m} \sim 1 \mbox{ TeV } 
\quad , \quad
\mbox{Im}(c^{qu}_{12}) \sim \frac{\alpha}{4\pi} \, .
\end{equation}
where $\tilde{m} $ is the typical scale of superpartner masses.
Notice that until last year this would have been more than welcome, as a possible signal of supersymmetric particles, while now it starts being in tension with the bounds from direct detection!

\spazio

For a detailed study along this direction we refer to \cite{Lodone:2012boh}, in which this possibility is thoroughly discussed.
It can be seen that the other bounds on flavor-violation in the quark sector are easily satisfied, while the Electric Dipole Moments (EDM) turn out to be the strongest constraint. This picture could also be extended to the lepton sector, and in this case the strongest bounds come from $\mu\rightarrow e\gamma$ and the electon EDM.

\section{Conclusions and outlook} \label{sect:concl}

We discussed the status of a few motivated beyond-the-MSSM models in the light of the recent LHC data.
Our conclusions can be summarized as follows:
\begin{enumerate}

\item {\bf Insisting on Naturalness: Natural Supersymmetry.} The present hint of a Higgs boson at 125 GeV tells us that, in the MSSM, the stop should be too heavy to be natural according to the criterion $\Delta \lesssim 10$. This instead is not a problem for non-minimal extensions, and thus we can at least say that beyond-the-MSSM models with increased quartic coupling are more motivated than before.
Apart from that, the most solid input so far is that the first two generations must be above the TeV, thus either one tolerates $\Delta\gtrsim 100$ or typically one needs the third generation to be lighter than the first two. For this reason, `More Minimal' or `Hierarchical' scenarios can now be renamed `Natural Supersymmetry'  (since it is the only left natural possibility... unless point 4, below).
Actually in $\lambda$SUSY the stop at about 1 TeV, eventually degenerate with the other squarks, can be compatible with $\Delta \lesssim 10$. In this case however also the first two generations can naturally be much heavier.
Low scale SUSY breaking is also favoured in this view, since broadly speaking it reduces the amount of tuning.

How to exclude Natural Supersymmetry? A relatively fair statement may be: \emph{not finding the Higgs boson} (which would actually exclude also most of the finely-tuned SUSY scenarios), \emph{and/or especially not finding superpartners at the LHC}.

\item {\bf Escaping the LHC bounds.} R-Parity violation after all is not compulsory, and without it the present bounds are partly escaped. Other possibilities can be SUSY with a compressed spectrum, Stealth mechanisms, or Dirac gauginos.
It is important to be ready to these distortions of the `usual' LHC phenomenology.

This is clearly a very hot topic! In the future, if the bounds get stronger, we might be forced to call it something like  `Alternative Natural Supersymmetry'...

\item {\bf Insisting only on DM and unification and/or strings: Split and High-scale SUSY.}
By construction, these possibilities are not touched by direct bounds on squarks. Thus for the moment the only experimental input for these models is the value of the Higgs boson mass, from which we can say that Split SUSY has to be `not-so-split' ($\tilde{m}\lesssim 10^5$ TeV), while nothing can be said so far in case of High-scale SUSY.

\item {\bf Higgs boson not at 125 GeV.} The importance of the issue is such that we cannot content ourselves with a $2\div3\sigma$ evidence: although there are a lot of good reasons for the Higgs boson to be there\footnote{See for example \cite{HiggsPredictions}!}, the significance of the present hint is such that it may well be a background fluctuation.
More importantly, heavier scalars with reduced couplings are present in SUSY as well as in other motivated extensions of the SM, and they must be looked for.

What can be said is that, if the present hint turns out to be a fluctuation, then it is obviously crucial to look for the most-SM-Higgs-like particle (with reduced couplings).
In this case, it is fair to say that such a particle cannot be considered to be reasonably excluded until the exclusion bound at 95\% c.l. is at least down to $\sigma / \sigma_{SM} \lesssim 0.1$.

\item {\bf Direct CPV in $D$ decays.}
In the context of Supersymmetry, assuming that the SM contribution is not enough to explain the data, a very natural possibility is to have `disoriented A-terms'.
This may be explicitly realized in a context in which the flavor structure has something to do with the mechanism of Partial Compositeness.
If this is the case, new effects are predicted to be around the corner in the neutron EDM and, if the picture is extended to leptons in the most straightforward way, in the electron EDM and in $\mu\rightarrow e\gamma$.

\end{enumerate}

\spazio
To conclude, in the near future a very fundamental question has a chance to be answered: \emph{was Naturalness a good guiding principle?} The LHC will tell us.
The very minimal requirement is a colored sector mainly coupled to the top, at or below the TeV. If Supersymmetry is the way of Nature, this will be the scalar partner of the top. Thus a very important point is whether the stop is there or not. Notice that if this sector is made of scalars the first option would be SUSY, while in case of fermions the most plausible possibility would be compositeness\footnote{For `historical' references see \cite{Kaplan:1983sm}-\cite{Dugan:1984hq}.}.

In any case, also being open to the possibility that Naturalness is implemented in an unexpected way, at least new particles/resonances within the LHC reach and/or a non-standard Higgs boson are requred.
Thus, in case the Higgs boson is discovered, it will be crucial to measure its couplings and check whether it is SM-like or not\footnote{See e.g. \cite{Falkowski:2012vh}-\cite{Giardino:2012ww} for recent studies.}, also looking at WW-scattering\footnote{See e.g.  \cite{Englert:2009av}\cite{Borel:2012by}\cite{Ballestrero:2012xa} and references therein.} and looking for a second Higgs doublet\footnote{See e.g. \cite{Blum:2012kn}\cite{Maiani:2012ij} for recent studies.}.
If the Higgs particle is discovered to be exactly SM-like and no new particles are found, this will be a \emph{strong indication that there is fundamental finetuning in Nature}\footnote{See e.g. \cite{Weinberg:1987dv} for another indication, as already said. In general, one can argue that finetuning with no symmetry explanation is an evidence for the Multiverse, or string theory Landscape. See also \cite{Giudice:2008bi},\cite{Agrawal:1997gf} for discussions.}, we like it or not.
On the contrary, a non-standard Higgs boson and especially new particles would probably tell us that naturalness arguments were correct.
Confirming or discarding the naturalness of the Fermi scale will be a very nontrivial conceptual input for the scientific though.
Thus for the moment it makes sense to keep insisting on Natural Supersymmetry, until it is found or excluded.

\begin{figure}[t]
\begin{center}
\includegraphics[width=1.0\textwidth]{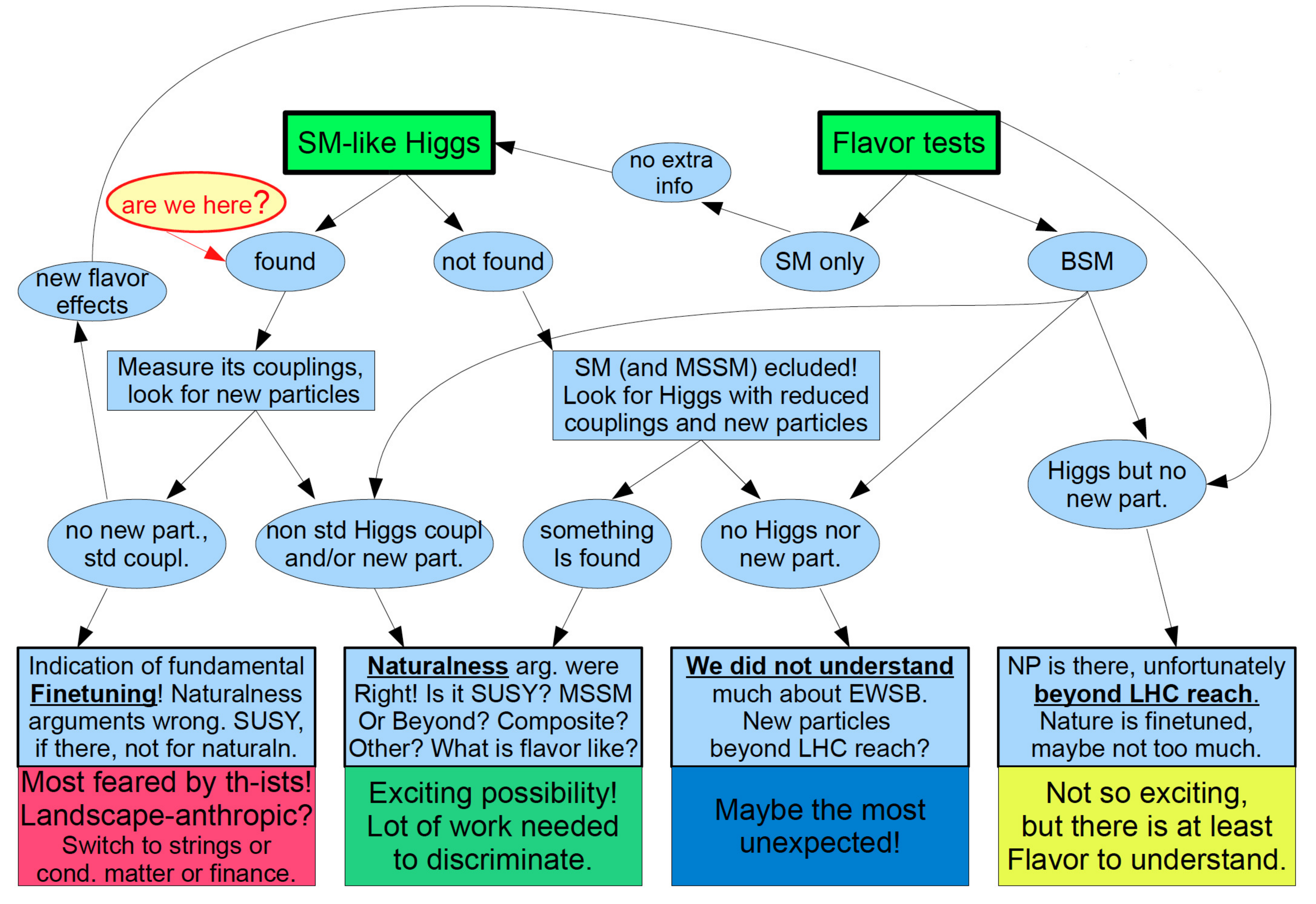} 
\caption{\small {\it General considerations about the LHC outcome. The important point is that the LHC will probe the concept of Naturalness. Thus it makes sense to keep insisting on Natural Supersymmetry, until it is found or excluded. What can be said in light of the LHC data so far is that the MSSM starts being a bit too finetuned if no extra ingredient is introduced.}}
\label{fig:outlook}
\end{center}
\end{figure}

A quite non-conventional alternative is that \emph{nothing} is found at the LHC, not even the Higgs particle, which is probably what most people expect least \footnote{One can think about configurations in which the lightest Higgs boson is completely `buried' under the QCD background, see e.g. 
\cite{Bellazzini:2009xt}\cite{Bellazzini:2010uk}. However there are typically other particles to be detected, such as the heavier scalars of the Higgs sector, and in conventional models it is unlikely that really nothing is found at the LHC.}.
Finally, we must not forget about Flavor Physics: even in presence of a totally-SM-like Higgs boson and no extra particles, non-CKM flavor effects would at least imply that there is some new physics relatively at hand, although maybe beyond the LHC reach for direct discovery.
Figure \ref{fig:outlook} is a pictorial summary of these last considerations.

\section*{Acknowledgments}

I thank Riccardo Barbieri, Gino Isidori, and Riccardo Rattazzi for useful discussions.
I also thank the organizers of the session ``Electroweak and searches'' at the ``XX International Workshop on Deep Inelastic Scattering and related subjects'' for their kind invitation to talk about this subject.
This work is supported by the Swiss National Science Foundation under grant 200021-125237.

\vspace{0.3cm}

\begin{multicols}{2}

\end{multicols}

\end{document}